\documentclass{article}

\usepackage{amsmath,amsfonts}
\usepackage{cite}

\usepackage{geometry,graphicx,url,caption,subcaption}
\usepackage[usenames,dvipsnames]{xcolor}
\definecolor{oxfordblue}{rgb}{0.0, 0.13, 0.28}
\definecolor{dodgerblue}{rgb}{0.12, 0.56, 1.0}

\usepackage{cancel}
\usepackage{ulem}

\begin{document}

\title{The Role of Network Analysis in Industrial and Applied Mathematics}

\begin{centering}
\author{
Mason A. Porter$^{1,2,3}$ and Sam D. Howison$^{2}$
\\
$^{1}$Department of Mathematics, 
UCLA, Los Angeles, California 90095, USA  \\
$^{2}$Mathematical Institute, University of Oxford, Oxford OX2 6GG, UK \\
$^{3}${CABDyN} Complexity Centre, University of Oxford, Oxford OX1 1HP, UK
} 
\end{centering}

\begin{abstract}

  Many problems in industry --- and in the social, natural, information, and
  medical sciences --- involve discrete data and benefit from approaches from
  subjects such as network science, information theory, optimization,
  probability, and statistics. The study of networks is concerned
  explicitly with connectivity between different entities, and it has become very
  prominent in industrial settings, an importance that has intensified amidst the modern data deluge. In this commentary, we discuss the role of network analysis in industrial and applied mathematics, and we give
  several examples of network science in industry. We focus, in particular, on discussing a physical-applied-mathematics approach to the study of networks. We also discuss several of our own collaborations with industry on projects in network analysis.
    
\end{abstract}


\maketitle



\section{Introduction}\label{sec1}

Mathematics has long played a vital role in industry. From the
mixing of fluids to produce an ideal bar of chocolate to the study of
gasoline emissions in vehicles, wonderful problems in continuous mathematics
--- typically framed in terms of ordinary and partial differential equations
--- have arisen from industrial problems~\cite{study_group}. They have
increasingly complemented the equally wonderful problems posed by
applications in fundamental science and engineering. The applied mathematics curricula
(and subjects studied by the academic staff) of many universities reflect this historical bias toward continuum problems.

The goal of the present commentary is to promote an approach to network analysis (especially in industry) through so-called `physical applied mathematics'. It is first useful to convey our perspective on such a viewpoint, which is one of the most prominent approaches to applied mathematics and our personally favored approach to science. In a physical-applied-mathematics approach to a problem, one uses basic physical (or biological or chemical)
principles and relevant domain knowledge to derive
governing equations (most often in the form of ordinary or partial
differential equations) and boundary and/or initial conditions; simplifies
the equations to make them mathematically tractable; studies the equations both computationally and with a wide
variety of mathematical tools (often approximately, such as with asymptotic analysis and perturbation theory); compares the numerical solutions (ideally of both the simplified governing equations and, if possible, the `original'
equations) with approximate analytical solutions or qualitative behavior
revealed through procedures like a dynamical-systems analysis; compares these
results with controlled experiments; and, when possible, compares the
experimental results with natural or industrial phenomena in more realistic
settings. Textbooks such as \cite{linsegal,andrewfowler} describe these
ideals. Through making comparisons, one also revisits one's assumptions, adjusts models, refines experiments, and so on. A final crucial step is to interpret the
results of the mathematical and numerical studies in a way that engages
seriously with the original problem. A problem's stakeholders must learn
something from the mathematical efforts, and such stakeholders --- whether they
are scientists in other academic departments, people who work in industry or government, or others --- often collaborate directly on the problem. At a minimum, they need to
be consulted early and often, as they offer domain knowledge.

Good physical applied mathematics can start from potentially any type of
problem, including fluid, solid, or granular phenomena in nature;
observations of biological systems; observations of human or animal behavior;
physical, behavioral, or other phenomena in industry; and much
more. Industrial problems (and other problems as well) typically start out in woolly
form, and a key challenge for applied mathematicians is formulating a
concrete, tractable mathematical problem whose solution (perhaps in
approximate or numerical form) can yield important insights about the
original problem or phenomenon. This is the art of
`mathematical modeling', and in industrial mathematics, 
it often entails taking a physical-applied-mathematics approach to problems that arise from
industry. It is \textit{{Applied}} mathematics, rather than applied \textit{{Mathematics}},
as it is crucial to engage very seriously with applications.

This approach, for which open-problem brainstorming workshops, known as `study groups', with industry have made
pioneering contributions \cite{study_group}, also applies to problems and
data that take discrete forms. It has long been true that many problems in
industry include discrete data and benefit from approaches that incorporate
topics from subjects such as network science, information theory,
optimization, probability, and statistics. For example, solving problems in
optimization is crucial for assembly lines, and the famous traveling
salesperson problem (TSP) has an undeniably practical origin \cite{tsp}. Amidst the modern data deluge, the importance of discrete data and associated approaches has reached new heights. Social media, which now pervade all aspects of life, involve interactions between entities; radio-frequency identification (RFID) device data track the
movements of people in cities and stores through discrete delineated zones; people have associated metadata
that describe their characteristics using discrete (categorical or ordinal)
variables; and so on. Network science is concerned explicitly with
connectivity between (and among) different entities \cite{newman2018}, and it has become very
prominent in industrial settings \cite{ejam2016} --- an importance that has been accentuated by the modern wealth of data. 

There is also a long history of using statistical approaches (such as actor-oriented models and others) to study networks \cite{faust1994,snijders2001,snijders2010}, and statistical approaches have long dominated industrial approaches to network analysis. When studying networks, however, it can be very fruitful to take a complementary perspective: one uses the established approach of physical applied mathematics, but now the problems need not be physical (or from other traditional domains), and in particular they are often discrete in nature and/or involve copious amounts of data. 
Specifically, as we discuss in more detail in Section \ref{sec2}, it is desirable to
combine network science, `the study of connectivity', with an applied-mathematics philosophy, which has been enormously successful in
collaborations with industry \cite{UKsuccess}.\footnote{See \cite{whisper} (from the collection \cite{UKsuccess}) for an example application of network analysis in industry.} For a recent collection of
modeling efforts involving networks (including in industry), see the December
2016 special issue of \textit{European Journal of Applied Mathematics}
\cite{ejam2016}. For some more specific examples, see the special issue of \textit{Royal Society Open Science} on urban analytics \cite{city2017}. There even exist companies that specialize in network analysis,\footnote{One example is Orgnet LLC \cite{orgnet}, with which neither of us has an affiliation or collaboration, which `provides software, training, consulting, and research in the application of network analysis in a wide variety of domains.' Essentially, they do network analysis and data visualization for hire. For instance, a company may hire them as a consultant to analyze their organizational structure as a network (say, of people connected based on their interactions with each other) to determine its key employees, including perhaps underappreciated people, for information flow.} and there are of course myriad companies that specialize in data analysis more generally. The latter category includes our collaborator dunnhumby \cite{dunn-blog}, and numerous other companies (including our collaborators HSBC, Tesco, and Unilever) include data and network analysis as part of larger research portfolios.

In this article, we discuss the role of network analysis in industrial and
applied mathematics, and we give several examples of our own work on network science in collaborations with
industrial partners. Specifically, we highlight a physical-applied-mathematics approach to industrial problems, though we are well aware that other perspectives are also important. In Section~\ref{sec2}, we discuss network modeling and relate it to
traditional ideas from physical applied mathematics. In Section \ref{sec3}, we
discuss applications of mathematics to the social sciences. In Section
\ref{sec4}, we discuss network science in industrial settings. We conclude in
Section \ref{conc}.


\section{Network Modeling}\label{sec2}


\subsection{Introduction and a Few Basic Ideas}

The study of networks incorporates tools from a wide variety of subjects ---
such as graph theory (of course),\footnote{By its nature, network analysis intersects significantly with graph theory, but the former has a much broader spectrum than the latter (e.g., dynamics often plays a central role), and graph-theoretic analysis is often not the primary focus in studies of networks.} computational linear algebra, dynamical
systems, optimization, statistical physics, probability, statistics, and more
--- and is important for applications in just about any area that one might
imagine. Scholars who study networks ask questions like the following: Who
are the most important people, and which are the most important collaborations, in a network of overlapping
committee memberships? What is a good movie-recommendation strategy in a
social network? How did ideas spread over Twitter and other social media in
the \#Brexit debate, and how did this spread of ideas influence opinions and
events? What role did the spread of misinformation, `fake news', and
`alternative facts' play in the 2016 US presidential election, and how did
this role change with evolving network structure as communication channels
became more polarized? How can one use information from social interactions
to improve strategies for predicting and preventing criminal activity? Which
parts of a granular material are the least stable, and how should one measure
this? How can one improve transportation systems and building layouts (e.g.,
the location of checkout tills and sale items in a supermarket) to mitigate
traffic congestion? How can one control cascading failures in infrastructure
or financial networks? How should one measure the robustness of power
grids to failures in transmission lines, and how should one design smart
grids to ensure robustness? How do financial assets coevolve, and what are
the best measurements to help forecast their future
evolution? How does the structure of an animal social network affect individual and collective behavior, and what types of interactions should one consider in such a network? 

To appreciate the use of a physical-applied-mathematical approach for the
study of networks, one first needs to have some idea of what a network is. In
its broadest form, a network is a representation of the connectivity patterns and connection strengths in a complex system of
interacting entities
\cite{newman2018, faust1994, kolac2009, jackson2010, stro01}.  Most
traditionally, a network is represented mathematically as a graph $G$, which
consists of a set $V$ of `nodes' (or `vertices') that encode entities and
a set $E \subseteq V \times V$ of `edges' (or `links' or `ties') that
encode the interactions between those entities. However, the term `network'
is more general than a graph, as a network can encompass connections among an
arbitrary number of entities, can have nodes and/or edges that change in
time, can include multiple types of edges, often have associated
dynamical processes both on the networks and of the networks, and so
on. Associated with a graph is an `adjacency matrix' ${\bf A}$, where, if one does not include a value to model the strength of a connection, an entry $a_{ij} = 1$ indicates the presence of an edge that connects entity $i$ to
entity $j$ directly, while $a_{ij}=0$ indicates its absence. 
That is, when $a_{ij}=1$, node $j$ is `adjacent' to node $i$, and the associated edge is `incident' from node $i$ and to node $j$. The number of edges that emanate from a node are its `out-degree', and the number of edges incident to a node are its `in-degree'. For an undirected network, $a_{ij}=a_{ji}$, and the number of edges incident to a node constitute the node's `degree'. The spectral properties of adjacency (and other) matrices give important information about associated graphs \cite{newman2018,piet-book}; for undirected networks, it is common to exploit the beneficent property that symmetric matrices only have real eigenvalues.

Although the study of networks continues to advance at a rapid pace, it can be useful to keep in mind some basic ideas. For example, it can be very insightful (e.g., for developing ranking methods for Web pages, sports teams, and other things \cite{gleich-sirev2015}), though sometimes fraught with complications, to study important (i.e., `central') nodes, edges, and other small structures in networks \cite{newman2018}. Another important theme is the study of both small and large mesoscale features, which impact network function and dynamics in important ways. Certain small subgraphs (called `motifs') may appear frequently in some networks \cite{milo2002}, providing possible indications of fundamental structures such as feedback loops and other building blocks of global behavior. There have been extensive studies of various types of larger-scale network structures, such as dense `communities' of nodes \cite{Porter2009, Fortunato2016} (see Section \ref{sec4}), core--periphery structure \cite{csermely2013}, and others. Other famous features of many networks have also played an important role in the emergence of `network science' as an area of study \cite{vesp2018}. One of these is the `small-world property' \cite{ws98,small-scholar}, in which the mean shortest-path distance between nodes in a network scales sufficiently slowly (specifically, logarithmically or slower) as a function of the number of nodes.\footnote{One has to define this property somewhat differently if one is studying a single, finite-size network rather than a sequence of networks of increasing size.} In many situations, such as in social networks, there is simultaneously significant local clustering. Another is heavy-tailed degree distributions (as idealized by a power law), indicating the presence of many nodes with a small degree but few nodes (sometimes called `hubs') with a large degree \cite{newman2018,ba99}.

To consider edges with different amounts of importance, one can assign a weight
--- typically a nonnegative real number, although there are many situations,
such as in the study of international relations or social interactions \cite{newman2018,faust1994}, in
which negative values can be appropriate --- so that the entry $w_{ij}$ of
a `weighted adjacency matrix' (or `weight matrix') ${\bf W}$ represents
the weight of the connections between nodes $i$ and $j$. A large value of $w_{ij}$ represents a strong connection from $j$ to $i$, though sometimes (e.g., for applications in transportation) one
might instead have a distance matrix, and then elements with smaller values
represent stronger connections. Such data can arise, for example, in the form
of physical distances (e.g., road networks) or from measurements of empirical or expected travel time between a pair of locations (e.g., using Oyster-card data from Transport for London). 


\subsection{Modeling Considerations and the Incorporation of Complications}

As we mentioned in Section \ref{sec1}, an important issue in the study of networks involves the notion of
`modeling' itself. An applied mathematical (or physical) model usually
takes as a starting point a postulated mechanism of cause and effect.
Statistical models, by contrast, have probabilities underlying them and are more (and sometimes purely) descriptive in nature.\footnote{Another approach in data science, such as in hierarchical clustering, is to `start from the data'.}
 Statistical models are often very successful at indicating correlations (interpreted broadly). Consequently, it is not surprising that the study of networks is more common among statisticians than among applied mathematicians, as reflected by the prominence of statistical approaches in studies of networks in industry and applications \cite{kolac2009}. However, statistical approaches are rightly cautious in deducing causation from correlation.  Hence, a distinctive
feature of network modeling in the spirit of physical applied mathematics is its linking of ideas
and tools from statistics (which are necessary, given the high-dimensional nature of networked systems) with the desire for causal
mechanisms. Put another way, a physical-applied-mathematics approach tends to put more emphasis on detailed modeling of mechanisms than do statistical perspectives. One possible desirable outcome is to derive some kind of
governing equations (perhaps high-dimensional ones), which one can try to
simplify using some sort of mean-field theory, master equation, or other
approximations \cite{newman2018,frontiers2016}. Unfortunately, this is often very
difficult.

Given data in which connectivity plays a role, and assuming that one wishes to
use tools from network science to help in one's analysis --- that itself is not
always obvious, and it is an important modeling decision --- one needs to decide what
type of network description to use, analogously to deciding a level of
description using other approaches (e.g., discrete particulate interactions versus a continuum (PDE) model). Just as one needs to use the correct conservation laws (and boundary conditions, initial conditions, and so on) in continuum models, it is crucial to choose an appropriate network representation for the problem at hand. If one studies the wrong network or asks questions that one's network representation cannot answer adequately, it is easy to end up with nonsense. In taking a physical-applied-mathematics approach, it is typically desirable (though it can be rather challenging to do it well) to proceed as follows: (1) propose mechanisms --- often probabilistic ones, such as interactions arising from a Poisson process \cite{Feller-book1} --- for the generation of edges and edge weights; and then (2) to interpret the ensuing results.

To be a bit more concrete, suppose that one possesses a
time-resolved data set representing social
interactions. The most common approach in such a situation is to aggregate
the data into a time-independent representation and study the resulting
graphs and adjacency matrices with standard tools, but that can cause several
problems. There are many choices for aggregation, the simplest of which is to
simply count the number of interactions between each pair of entities and
place those numbers in the weight matrix ${\bf W}$. If, for example, $i$ and $j$ interacted with each other twice during the monitored time
window, then $w_{ij} = w_{ji} = 2$. Unfortunately, this type of aggregation ignores the `bursty' nature
of dynamics in social systems and instead is based on an implicit (and often incorrect) assumption of
Poissonian-in-time interactions between entities \cite{till2012}. One can try to
aggregate the temporal information in a more sophisticated way, but then one
has to think very carefully both about the observations and about the sociological (or
other) model of communication between individuals. Such aggregations of
temporal data into time-independent representations also suffer problems
related to concurrency and ordering of interactions, which is crucial for
applications such as transmission of information and diseases (and thus for
many problems of industrial interest), so one needs to go beyond the
traditional tools associated with time-independent graphs. One way to do this is to study
`temporal networks' \cite{Holme2012,Holme2015} and either perform
aggregations over multiple windows (which can either overlap or not) or
perhaps not aggregate at all and consider the timeline of interactions to be
the objects of interest. Indeed, the study of temporal networks is a
very active area of network science, with several actively researched, unresolved theoretical issues:
\begin{enumerate}
\item{It is very far from clear how to generalize measures and approaches (e.g., `centrality' measures
of node and edge importance, data clustering methods, and others) from
time-independent networks to temporal networks.}  
\item{These concepts can be generalized in many different ways, and which
generalizations are better for which applications, problems, and data is an open issue.} 
\item{One has to consider the important issues of discrete versus continuous time
and of interaction duration.} 
\item{One must also consider the relative
temporal scales of changes in network structure (weights and/or connections)
and changes in the states of network nodes and edges (e.g., time-dependent
traffic on city streets).} 
\end{enumerate}
Thus, with temporal networks (and any other generalization of ordinary
graphs), there is a modeling tradeoff: Should one collapse the data and use a
simpler representation, possibly losing something vital or even obtaining a
qualitatively incorrect answer in the process, to be able to use a
better-developed and better-understood approach; or should one keep some of
the salient information --- surely one cannot keep all of it, given
limitations imposed by data size and measurement --- and have to generalize the mathematical approach and perhaps make some
missteps along the way? As with mathematical modeling (and very prominently
indeed in industry), which approach to take depends on the problem and the
question that one is asking. Ideally, one pursues both approaches, because it is necessary
develop a better understanding of which simplifications are acceptable.

Temporal dynamics is not the only type of complication in interaction data. For example, data can have
`multilayer' structures, perhaps through the interaction of multiple
subsystems or through the presence of multiple types of connections (e.g., there can be multiple communication channels or multiple modes of transportation) \cite{Kivela2014,Boccaletti2014}, and one thus has to consider whether to use
a monolayer (i.e., single-layer) or multilayer approach. As with temporal networks, one keeps more
information with a multilayer approach, but it is challenging to generalize monolayer measures and methods, as different
generalizations are appropriate in different situations. There are several
other similar issues. Should one consider just network structure, a
dynamical process on a network, or an `adaptive network' \cite{thilo-adaptive}, in
which the dynamics on top of networks are coupled to the dynamics of the
network structure (e.g., a driver changes his/her route based on traffic
conditions)? Should one include annotations (e.g., demographic data) on nodes
and/or edges? Should one allow `hyperedges' or simplices to connect more than two nodes at a time? Should one perhaps do all of these things (as well as others that
we have not mentioned)? However, including everything yields an enormous mess that nobody knows how to study!


\subsection{Networks and Data}

Further issues arise in the form and fidelity of data. Data may be inaccurate
or missing (and `Big Data'\footnote{It often feels like many people believe in the Data Fairy. Perhaps they put their hard drive under their pillow and hope that the Data Fairy comes during the night to leave them a clue?} is very far from the same thing as `good data'
or `appropriate data'), and generalizing network structure to incorporate
more features necessitates demanding reasonable measurements of more
things. Thus, what data one can reliably collect (or obtain access to) will
also influence the complexity of the chosen network representation. There is
also the issue that most data do not come initially in the form of a network,
or it may come in such a form but with difficulty in determining the weights
(or even existence) of edges between entities.\footnote{As Tom Petty and the
  Heartbreakers might sing, the weighting is the hardest part.} In some
situations (e.g., physical networks, such as in the study of traffic on road
networks), there are straightforward ways to measure edge weights. However, in many other examples (e.g., from social or biological interactions), it is much
harder to reliably calculate the weight of a network edge from empirical
data. For example, suppose that data arises in the form of pairwise
similarities between entities. One can construct an adjacency matrix and thus
a network, but is a network approach the best one (or even a good one) to
use? Perhaps one should instead use data-reduction techniques from machine
learning? To give an even more complicated scenario, one may possess coupled time series, so one
can construct a (possibly time-dependent) set of similarities using one or
more of many possible ways of measuring similarities between time series, and
one thereby obtains a network (either a temporal one or time-independent one)
to analyze. However, one started with a set of coupled time series, so maybe one should use
time-series approaches?

Another salient point, which Andrew Stuart has pointed out
\cite{stuart2017} in the context of data assimilation (see \cite{law2015} for
an introduction to data assimilation) and which we borrow for our
discussion, is the level of verifiability (a kind of trust) of models in
different domains and how that affects mathematical (and statistical) modeling and the
interaction between data and models. At one extreme 
--- i.e., most verifiable --- lie physical models, in which there is a set of mechanistic equations, which, in many
cases, one trusts fully because they are derived from fundamental physical
principles that are supported by numerous repeatable 
experiments with very precise
and accurate results.  Somewhat less
verifiable (or calibratable)
are typical models from biology, which are often exploratory
phenomenological models. Nevertheless, there is still often a strong level of
support and/or confidence in them, and they often incorporate causal mechanisms based
on observation. Models from mathematical finance, as
used by financial institutions, often also lie at this level. They can be
well-calibrated to data, and they have a built-in causal mechanism
(e.g., no arbitrage), but they also have some inbuilt ad hoc assumptions. Even trickier (in the context of
verifiability) are many models in the social sciences, which are often based on sociological theory and/or thought experiments with much less direct observational support than the
ones in biology or mathematical finance.
Their role is often to illustrate or probe a postulated
mechanism or process embedded in a larger and more complex context,
but with little (or no) expectation of direct comparison
with data.  (See Section \ref{sec3} for examples and a discussion of mathematical modeling in the social sciences.)
Some of these models also rely on assumptions, such as perfect rationality,
whose validity is hotly contested. Finally, there are models, in fields
such as commerce and many others, where one usually possesses only empirical data and simple
agent-based or machine-learning models, but one aims to tackle large-scale and complex
problems by simulation. 
An important question is the following: How does the
interaction between modeling and data differ in these different scenarios? In
a physical model, in which one believes (at least approximately) the
governing equations, data may simply yield initial conditions, boundary
conditions, and parameter values. Naturally, data also constrains statistical models. At the extreme end of the spectrum, for many
complex systems, one does not have equations, and a model may be entirely
data-driven or even purely descriptive.\footnote{See \cite{kutz2013} for additional
discussion of data-driven modeling in complex systems (for scenarios with
various levels of trust), and see \cite{oreskes1994} for a discussion of model verification and validation in the context of the earth sciences.} It is important to emphasize that the location of a
particular model on this spectrum depends strongly on the availability of
widely-accepted basic principles, but it is also true that each
discipline has its own prejudices in favor of some kinds of models and
against others.


\subsection{Taking Tools Off the Shelf versus Designing New Tools (and New Shelves)}

A key challenge in network science --- and a major difference with other areas
of industrial and applied mathematics --- is that the off-the-shelf tools are
much less developed in network science than they are in other areas. It is
far from clear how to generalize tools and methods from graphs to more
complicated types of networks (and there remain a wealth of open problems even
when studying graphs), and such efforts are perhaps the most active part of network
science. When faced with a problem from industry (or elsewhere), there is a
tension between (1) simplifying it and using available approaches and
(2) trying to develop the new mathematical and computational tools that are
necessary to examine the problem in a more detailed and perhaps more
appropriate way.


\section{Applying Mathematics to the Social Sciences}\label{sec3}

An area that helps set the stage for the importance of networks in industrial problems is the application of mathematics, and networks especially, in the social sciences. The use of mathematical approaches to social phenomena
is much older than widely appreciated
\cite{faust1994, kochen1978}. More recently, the wealth of social data --- e.g.,
from social media such as Twitter and Faceook, and directly from companies in
forms such as mobile-phone data, shopping data, human movement data, and
others --- has brought social science to the center of the `Big Data'
explosion. In contrast to much smaller data sets, collected traditionally in forms
such as surveys, the data deluge has led to the formation of subjects such as
`computational social science' \cite{science2009, salganik}. This has placed
social science in a transition period in which an increasing number of
researchers who are trained in subjects such as computer science, physics,
and mathematics are trying to apply their techniques to social systems
\cite{watts2017}. There is an ongoing revolution in our ability to predict
and explain their dynamics \cite{hofman2017}, underscoring the importance of developing new mechanic models for use in computational social science \cite{holme2015-ss}.


\subsection{Example: Influence and Opinions on Networks}

To give a concrete example of mathematical modeling in the social sciences,
we discuss ideas related to opinions and social influence on networks
\cite{lehman-ahn-book, frontiers2016, loreto2009}. One of the best-known, albeit in many ways
rather naive, model for social influence on networks is the Watts Threshold
Model (WTM) \cite{watts2002}, which models the adoption of a product, idea,
innovation, or meme in a social network. Independently, the WTM was studied
slightly earlier by Valente \cite{valente95}, who incorporated network
structure into an influence update rule that Granovetter proposed in the
1970s \cite{granovetter1978}. The WTM is also a generalization of a process
known as `bootstrap percolation' \cite{bootstrap}. In the WTM model, and
other models of social influence, it is interesting to ask if an idea takes root
(e.g., if a product `goes viral' and reaches a large fraction of
individuals in a network), how long it takes to take root, where in a
network it is optimal to seed a new product, and so on. There are also
numerous other types of opinion models, often in the form of `voter
models', in which scholars examine whether consensus occurs, how long it takes to
occur, how many groups of different opinions are present at
steady state, and so on \cite{frontiers2016, loreto2009, lehman-ahn-book}. These results depend
both on network structure and on the rules that govern how node (or edge)
states are updated, and different qualitative results can arise from rather
specific, seemingly-minor choices (e.g., drawing an edge uniformly at random
in a voter model versus drawing a node uniformly at random in a voter model
and then picking one of its incident edges) \cite{frontiers2016, masuda2016, redner2017, sood2008}. The fact that such choices can have drastic effects on qualitative dynamics is a major modeling issue, especially if one starts with
an ill-defined, woolly problem, as is almost invariably the case when
collaborating with industry.

Suppose that one has a network, which can arise either from empirical data or
from a synthetic construction (perhaps as an instantiation of a random-graph
model). For simplicity, let's suppose that this network is represented by a
graph and is both unweighted and undirected. The WTM is a binary-state
`threshold model' \cite{gleeson2013,lehman-ahn-book}. In such models, each node $i$ has a
threshold $R_i$ (which, in most models, is time-independent) that is drawn
from some distribution. At any given time, each node can be in one of two
states: $0$ (inactive, not adopted, not infected, etc.) or $1$ (active,
adopted, infected, etc.).  The states of the nodes change in time according
to an update rule, and one can update nodes either synchronously or
asynchronously. In the context of spreading of information, one can construe the transition from $0$ to $1$ as an instantaneous purchase of a product or as representing --- in an extreme simplification --- the instant of a recognizable change in opinion. When updating the state of a node in the WTM, one compares the node's fraction $m_i/k_i$ of infected neighbors (where
$m_i$ is the number of infected neighbors and $k_i$ is the degree of node
$i$) to the node's threshold $R_i$. If node $i$ is inactive, it then becomes
active (i.e., it switches to state $1$) if $m_i/k_i \geq R_i$; otherwise, its
state remains unchanged. One can think of the quantity $m_i/k_i$ as a peer
pressure, and one can think of $R_i$ as representing stubbornness or inertia. One way to generalize the WTM is by calculating peer pressure in different ways. For example, in a model introduced in \cite{melnik2013}, there are three states:
nodes can be passive, active, or `hyper-active', where the last category of
nodes, which could represent leaders in a mass movement, exert more
influence than nodes that are merely active.

Threshold models are rather simplistic, and it is natural to ask whether there
exist real-life scenarios in which such models are appropriate for explaining
empirical observations \cite{frontiers2016, FB-profile-pic}. Although a binary
decision process on a network is a gross oversimplification of reality, it
can already capture two very important features \cite{oliver1985}:
interdependence (an entity's behavior depends on the behavior of other
entities) and heterogeneity (differences in behavior are reflected in the
distribution of thresholds). Typically, some seed fraction $\rho(0)$ of nodes
is assigned to the active state, although that is not always true (e.g., when
$R_i < 0$ for some nodes $i$).  Depending on the problem under study, one can
choose the initially active nodes either randomly (often uniformly at random) or deterministically.  For the
latter, for example, one can imagine planting a rumor at specific nodes in
a network, or perhaps this is a seed node that is trying to spread
misinformation, `fake news', or `alternative facts'.

There are many commercial and governmental applications of social influence
and opinion dynamics on networks. An effective model of a marketing campaign ---
whether to promote a product (or an idea or desired behavior, such as a
healthy diet), to perform targeted and personalized advertising, or to
counteract the spread of misinformation --- requires one to understand how
network structure can affect network dynamics and how different
dynamical processes (even ones that cosmetically seem rather similar to each
other) can exhibit qualitatively different behaviors. These are active
research areas with numerous fascinating theoretical issues (in mathematics,
social science, human behavior, economics, and more), methodological issues
(for both analytical theories and computation), and commercial, governmental,
and societal applications. In the commercial sector, for example, research on
human opinion, behavior, and influence can play a significant role in
personalized coupons. If, as sociological theory suggests, `you are who your
friends are' (as social network structure arises in large part from
homophily) \cite{faust1994}, one can imagine that coupon-program design
should include information about what one's neighbors in a social network are
buying.


\subsection{Ethical Considerations}

Mathematical and computational scientists are faced increasingly with ethical issues in their data analysis and other research \cite{data-theme,loukides2018,loukides2018-collection}. These issues --- which are diverse, multifaceted, and interdependent --- include replicability \cite{stodden2018}, accessibility of code and data, tension between replicability (and availability) and privacy of human data, and others. In the context of network analysis in industry, we would like to briefly discuss data ethics. For a brief introduction and pointers to many references, see the slide deck \cite{mason-slides}, a recent theme issue \cite{data-theme} of a {\it Royal Society} journal, and a recent conference \cite{conf} at University of Cambridge on ethics in mathematics.

Privacy is one of several crucial ethical issues raised by the analysis of
personal data \cite{mason-slides,grindrod2016}. Because data often include either explicit or implicit information about interactions between entities, one can use network analysis to de-anonymize data, especially when there is data about many of the same people in
different social networks \cite{anon1,anon2}. This becomes especially salient in light of the fact that personal data (e.g.,
medical records) can influence insurance premiums or other important items. The privacy breaches via Facebook of the defunct company Cambridge Analytica are now infamous (see, e.g., \cite{motherboard}), and these issues are also paramount when analyzing data from social media for research purposes \cite{emotions-guardian}. Additionally, although privacy may be the most obvious ethical dilemma facing researchers who analyze personal data, other ethical issues are also present, and applied mathematicians may encounter them in their research. 

Most mathematical scientists have insufficient ethical
training for the era of Big Data, and it is necessary that such training
be built into their education. Much of the social data used in collaborations
with industry raise these issues rather prominently, so this will become
an increasingly important aspect of industrial mathematics with social data
--- and especially with network data, as the ties between people can play a
major role in invasion of privacy and removal of anonymity. Several years
ago, we (MAP, with help from SDH) set up a procedure for studying human data
for the Mathematical Institute at University of Oxford (the current version
is available at \cite{MI-ox-data}), and we would like to see this kind of
approach becoming standard.


\section{Connecting with Industry}\label{sec4}

To illustrate the role of network analysis in industry through the lens of physical applied mathematics, we now briefly discuss a few of our past and ongoing projects. During the last decade, we have cosupervised several doctoral students jointly with industrial partners, initially through EPSRC Collaborative (Industrial CASE) awards and more recently through programs such as University of Oxford's Center for Doctoral Training (CDT) in Industrially Focused Mathematical Modelling (InFoMM)
\cite{infomm}. All together, this covers half a dozen students: four in collaboration with the
investment bank HSBC, one with the supermarket company Tesco, and one with the customer-science company dunnhumby. 

Our collaboration on network science started with our students who worked on
problems in partnership with HSBC, and (\textit{very importantly}) it has led
directly both to results of interest to stakeholders (HSBC and their clients)
and to us and the applied mathematics and network-science communities more
broadly. As these projects illustrate, the applications and the mathematics
are intertwined, as each drives the other in a crucial way. Early work on time-dependent correlations of financial assets
\cite{fenn-pre, fenn-chaos, fenn-qf} included ideas from network science and
random-matrix theory \cite{fenn-pre} and led to HSBC's `risk-on, risk-off'
(RORO) analysis of market behavior. RORO has been featured prominently in financial circles
(including as a tag in the \textit{Financial Times} blog); see, e.g.,
\cite{roro1}. 

Our work with HSBC helped pave the way for our more
recent work on financial assets and multilayer networks and our current
project on consumer--product purchasing networks. In parallel, it
stimulated theoretical analysis of time-dependent networks. 
The starting point was the study of `community
detection' \cite{Porter2009, Fortunato2016}, an approach to network
clustering in which, in some hopefully optimal way, one seeks to
algorithmically find dense sets of nodes that are connected sparsely to other
dense sets of nodes. In the HSBC-related work in
\cite{fenn-chaos, fenn-qf}, led by our doctoral student Dan Fenn, we detected communities by optimizing a `modularity' objective function
\cite{newman2018,newman2006pre}. We used a measure of
similarities in the time series of the exchange rates, based on a time-window aggregation, to compare observed connections in a
network with `random ones' in a null network\footnote{We primarily use null models, and null networks that are generated from null models, as structures to compare to those from empirical data. This resembles the standard use of null models in statistics, but it is not exactly the same.} constructed from a
random-graph model. We did this separately in each
window, but we connected the windows sequentially, with a temporal overlap between consecutive windows.  

Informed by this work --- and also work in other applications, such
as legislative cosponsorship and voting networks \cite{zhang2008, waugh2009} --- MAP and
collaborators developed a more principled approach to community detection in
time-dependent networks with discrete temporal `layers' that can represent
interactions at one time point or over some period of activity
\cite{Mucha2010}. In this approach, one incorporates contiguity between layers using `interlayer' edges between nodes in different
layers. Mucha et al. \cite{Mucha2010} derived a generalization of the
modularity objective function for the resulting `multilayer' network
\cite{Kivela2014}, and they maximized it to assign node-layer pairs to
communities. Thus, an entity can be assigned to
different communities in different time periods, and one can study the evolution of community structure over time. One can also use multilayer modularity for other situations, such as `multiplex' networks, in which
there exist multiple different types of edges.

The work of~\cite{Mucha2010} left open numerous questions about multilayer
networks in general and about community detection in such networks in
particular. Our student Marya Bazzi, also funded by a CASE award in
collaboration with HSBC, revisited some of the applications and data sets
studied earlier by Dan Fenn using the more sophisticated
multilayer-network techniques, which had been developed and advanced during
that time, to examine those phenomena in a more sophisticated manner
\cite{Bazzi2014}.\footnote{Following the 2010 publication of \cite{Mucha2010}, the study of multilayer networks has become one of the most prominent areas of network science \cite{Kivela2014,Boccaletti2014}; and several papers, such as \cite{Peixoto2015,DeDomenico2014a} and others, have proposed different approaches for studying communities and other mesoscale structures in such networks.} In addition to affirming the results (such as structural changes in the networks following the Lehman
Brothers bankruptcy in 2008) from the work of Fenn et
al. \cite{fenn-pre, fenn-chaos, fenn-qf}, Marya also found subtler structural changes that
merit further exploration.  As the focus of the work shifted from application to theory, she made several
theoretical and methodological advances, first in \cite{Bazzi2014}
and then (jointly with us, fellow University of Oxford doctoral student Lucas Jeub, and our
collaborator Alex Arenas) in \cite{bazzi2016bench}. Advancements in
\cite{Bazzi2014} include the careful distinction between null models and null
networks in modularity maximization, proofs of crucial conceptual ideas in
multilayer networks (e.g., that the limit of zero interlayer coupling is a
singular one), and toy examples that set the stage for our recent systematic
development of flexible generative models for multilayer networks in
\cite{bazzi2016bench}. These models allow a wide variety of correlations
across different layers --- a feature that is very important for real multilayer
networks.  

Our current work in collaboration with supermarket and customer-science companies partly builds on
the above insights on clustering in networks and partly moves in entirely new directions. In one project, we are clustering shopping data
and trying to incorporate constraints and metadata that are appropriate for
that application. In another, we are developing a generative model for shopping trajectories of people in supermarkets, with the hope of finding good approaches for relieving congestion.

\begin{figure}
\begin{center}
\includegraphics[width = 0.6\textwidth]{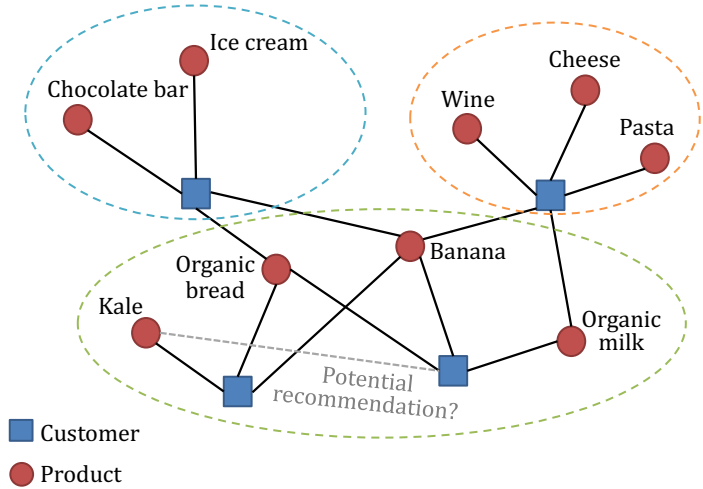}
\end{center}
\caption{A schematic bipartite (i.e., two-mode) network in which 
consumers are adjacent to the products that they purchase. 
The blue squares are consumers, and the red circles are the products. The dashed ellipses enclose nodes 
that have been assigned to the same community using a clustering 
technique. The gray dashed edge towards the bottom of the figure is a 
potential recommendation that one might give to a consumer, as 
there is a node that is not yet purchasing kale but who has been 
assigned to the same community as kale. 
[This figure is a slight modification of one that was created by Roxana Pamfil.]
\label{bipartite}
}
\end{figure}

The project of our doctoral student Roxana Pamfil lies within this strand of
research but also considers a new application, whose structural constraints differ in important
ways from the ones in the above problems. Our work with Roxana \cite{roxana-infomm} is in collaboration with dunnhumby, a customer-science company. She is analyzing data from anonymized consumers and the products that they purchase in their shopping baskets (from many Tesco stores in the United Kingdom). From these data, we construct bipartite (i.e., two-mode)
networks in which consumers are adjacent to purchased products. See Fig.~\ref{bipartite} for a schematic. The bipartite structure
needs to be incorporated into methods for clustering the data. Determining the edge weights also requires considerable
care. For example, one can use so-called `item
penetration' (the fraction of all of the items bought by customer $c$ that were
product $p$), `basket penetration' (the fraction of all baskets of customer
$c$ that included product $p$), or something else. Different choices can yield
qualitatively different results, and a key challenge is to determine
precisely which weights are most appropriate for which questions and which of
them give the most robust results. 
   
\begin{figure}[htbp]
\vspace{.5cm}
    \begin{subfigure}[b]{0.31\textwidth}
      \includegraphics[width=\textwidth]{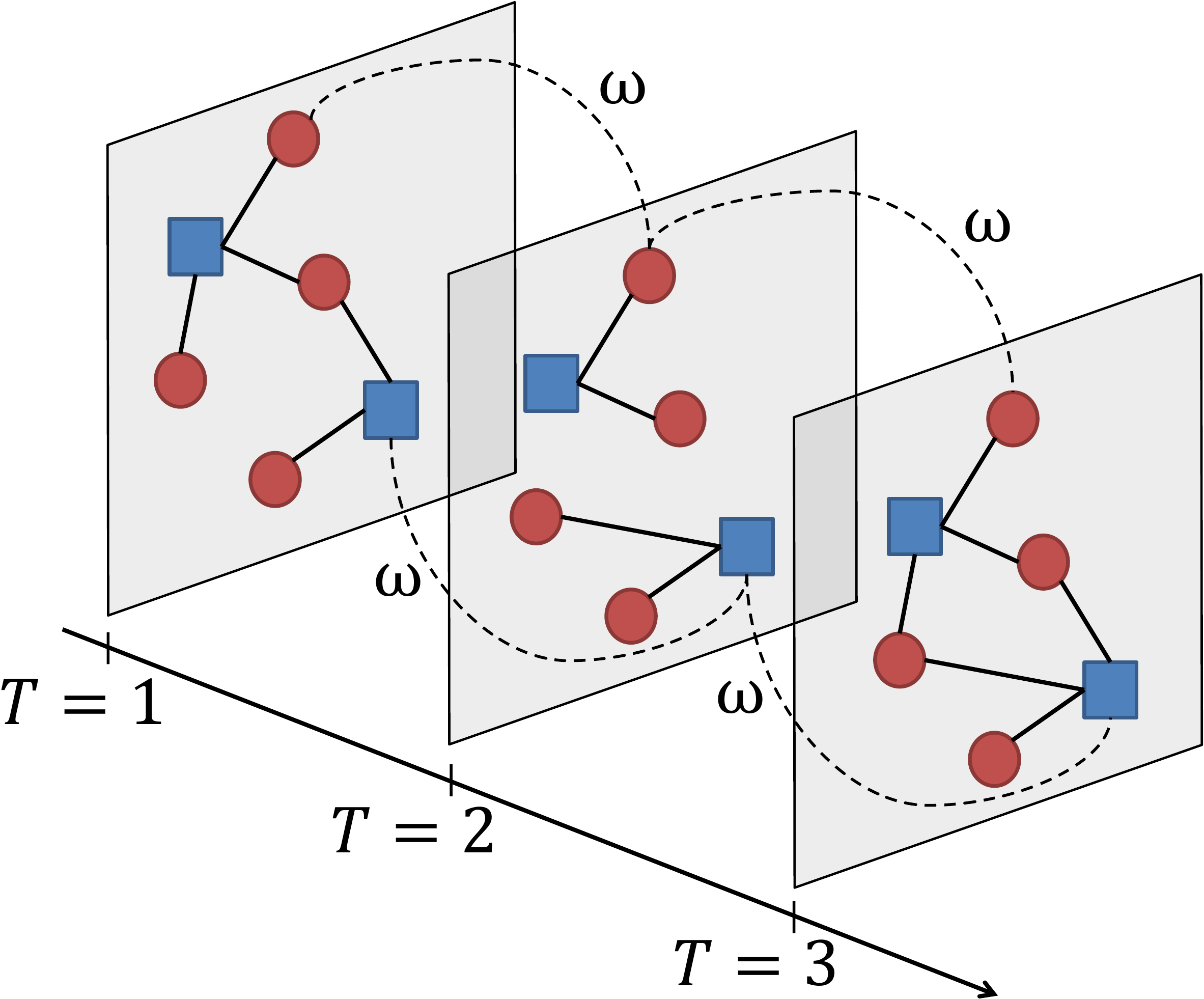} \smallskip
      \caption{Time-dependent networks}
    \end{subfigure}
    \hfill
    \begin{subfigure}[b]{0.31\textwidth}
      \includegraphics[width=\textwidth]{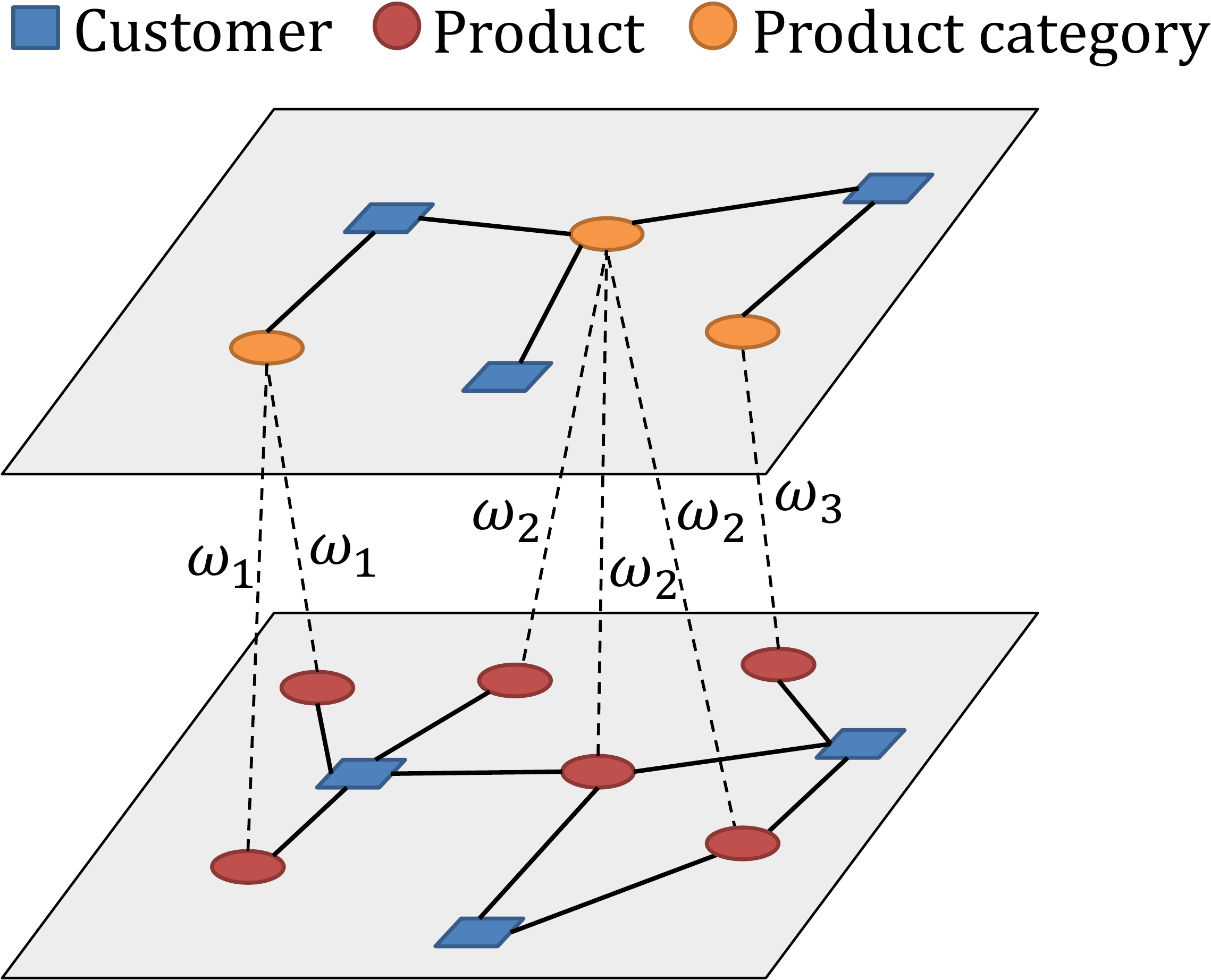} \smallskip
      \caption{Multilevel networks}
    \end{subfigure}
    \hfill
    \begin{subfigure}[b]{0.31\textwidth}
     \includegraphics[width=\textwidth]{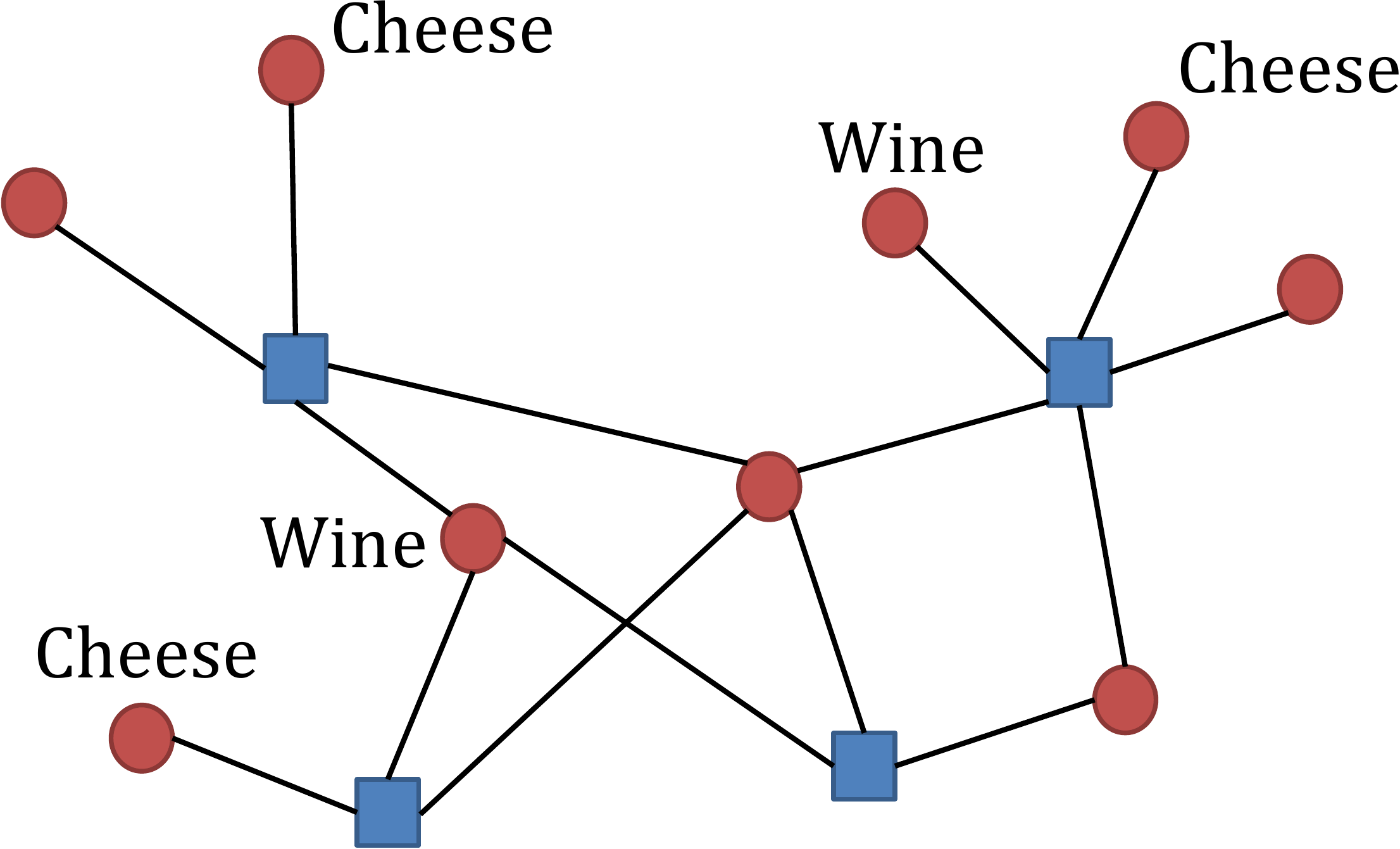} \smallskip
     \caption{Annotated networks}
    \end{subfigure}
  \caption{Schematic of network structures that are relevant for 
our work on product--purchase networks. (a) Time-dependent network, 
encoded with a multilayer representation, with time layers that indicate 
purchases by consumers (blue squares) of products (red circles) in 
supermarkets. The interlayer edge weights $\omega$ encode dependencies 
across time layers. Determining values for $\omega$, ideally from data or 
as an output of analysis, is an open problem in the study of 
multilayer networks. 
(b) A multilevel network, in which different layers represent 
different hierarchical levels in product descriptions. Note that the 
interlayer edge weights $\omega_i$ can be heterogeneous, as is also 
the case for multilayer networks more generally. 
(c) Annotated networks, in which we use product categories as product-node labels. 
[This figure is a modification of one that was created originally by Roxana Pamfil.]}
\label{complicated}
  \end{figure}

 We have access to time-resolved data, which are collected from several different stores (including multiple different store formats) and which include various shopper metadata. We also have access to product descriptions at different hierarchical `levels' (e.g.,
organic milk versus a particular type of organic milk), opening the door for
multilevel modeling with interlayer edges that represent inclusion
relationships and induced intralayer edges, whose existence can be inferred on
a different layer of a multilayer network. For example, if a consumer bought
a particular type of organic milk, then he/she necessarily purchased organic
milk more generally. For examples of networks with various complications, see
Fig.~\ref{complicated}.

We have also been incorporating statistical thinking into our collaboration with dunnhumby. An important aspect of our project with dunnhumby is clustering with mesoscale structures other than the `assortative' ones (which correspond to adjacency matrices that
look dense in the main block diagonal) that are typified by traditional community
structure \cite{Fortunato2016}. We do this in a statistically
principled way by using `stochastic block models' (SBMs), in which one specifies
a block structure and tries to find the clustering that best fits that
structure \cite{Fortunato2016,peixoto2017}. In this approach, statistical
inference takes center stage, and statistical model selection --- e.g.,
between different block structures or between whether or not one allows
overlapping communities --- becomes a key consideration
\cite{peixoto2017,Peixoto2015b}. In a multilayer setting, as we showed in
\cite{bazzi2016bench}, one can incorporate interlayer dependencies directly
into generative network models (such as SBMs) in a
convenient way, and we are actively using these ideas for clustering in
consumer--product purchasing networks. Roxana's work on SBM inference includes studying annotated networks (see Fig.~\ref{complicated}c), in which we incorporate node labels (e.g., consumer type). This is helpful for making recommendations for a newly-introduced product, for which no network data are available. Specifically, one can use the annotations and the inferred relationship between annotations and mesoscale structure to assign this new product to a community, which in turn indicates relevant consumers for that product.

As with our other collaborations with industry, Roxana's work has also led to new mathematical results, such as extensions to weighted bipartite networks of methodology from \cite{newman-clauset2016} for detecting communities in annotated networks and improved methodology for how to determine interlayer edge weights in multilayer SBMs in a principled way \cite{roxana2018}. The latter is a significant extension of Newman's recent work \cite{newman2016pre} that established an equivalence between modularity maximization (an ad hoc approach to community detection \cite{Fortunato2016}), with a principled choice of a resolution parameter, and a special case of an SBM in monolayer networks. Roxana's results are also important in network science more broadly, as most work on multilayer networks still use ad hoc weights for interlayer edge weights.

Roxana's project is a good illustration of
the way in which model choices, analysis, and simulation interact with data
in an iterative and question-driven way.
A crucial point to stress once again is the benefit
to both industrial and academic partners. In this project, the information in
the data and the structural constraints of the application yield multilayer
networks with different structures and metadata (and different sparsity patterns in the
network connections) than what has been analyzed previously. To further
develop the theory of multilayer networks, which is one of our primary
scientific interests and is one of the most active areas in network
science, it is necessary to consider diverse structures, applications, and ensuing
challenges. Otherwise, one risks developing a biased theory that hasn't been
tested adequately on relevant structures. For our industrial partners,
Roxana's project helps improve understanding of different types
of edge weightings between consumers and the products that they purchase, how
to categorize different types of customers, and the development of strategies
for product recommendations and personalized coupons (through reward cards,
which are also helpful for gathering data). Eventually, it is also desirable to account for
geographic variabilities in how people shop and for large-scale changes in
customer preferences over time. Long term, it is also valuable to combine these
insights with social-media data (e.g., with recommendations that are also
influenced by friends' purchases), though that will of course involve very
serious ethical considerations regarding what research in that direction is appropriate.

We hope to be able to use our models and data analysis in a predictive way to make product recommendations by assigning probabilities to unobserved edges (so-called `edge prediction' \cite{link2017}). Here, too, one faces a choice of model and level of sophistication: Should one use monolayer or multilayer networks, an SBM or modularity maximization, unweighted or weighted edges (for various choices of weights), and so on? As part of her thesis, Roxana has also modeled edge correlations in multilayer networks, and this too entails choices. However, because people tend to buy the same things over time, modeling correlations explicitly should improve edge-prediction results, and this is important for our work with dunnhumby and for many other applications. Excitingly, some of Roxana's work has suggested `experimental' work with dunnhumby to further evaluate her insights from methodological development, modeling, and data analysis.

Our latest doctoral student, Fabian Ying, is working on a project in collaboration with Tesco \cite{fabian-infomm}, which is also in collaboration with our colleague Mariano Beguerisse D\'iaz. Fabian is also examining data from Tesco, though his project --- investigating human mobility and congestion inside supermarkets --- is rather different from the ones that we
described above. The study of human and animal mobility is one of the most fascinating areas in complex systems \cite{mobility-review}, and investigating human mobility in supermarkets --- which occurs on shorter time scales and smaller spatial scales than almost all existing studies of human mobility --- is important for several questions (see \cite{riefer2017} for a recent study), including the following: How do customers shop and navigate within a supermarket? What is the best store layout to
reduce congestion? Where in the stores should the promotional items be
placed? For our industrial partner, these questions are of course related to
one of their major questions: How do they maximize revenue?

\begin{figure}
(a) \includegraphics[width = 1.0\textwidth]{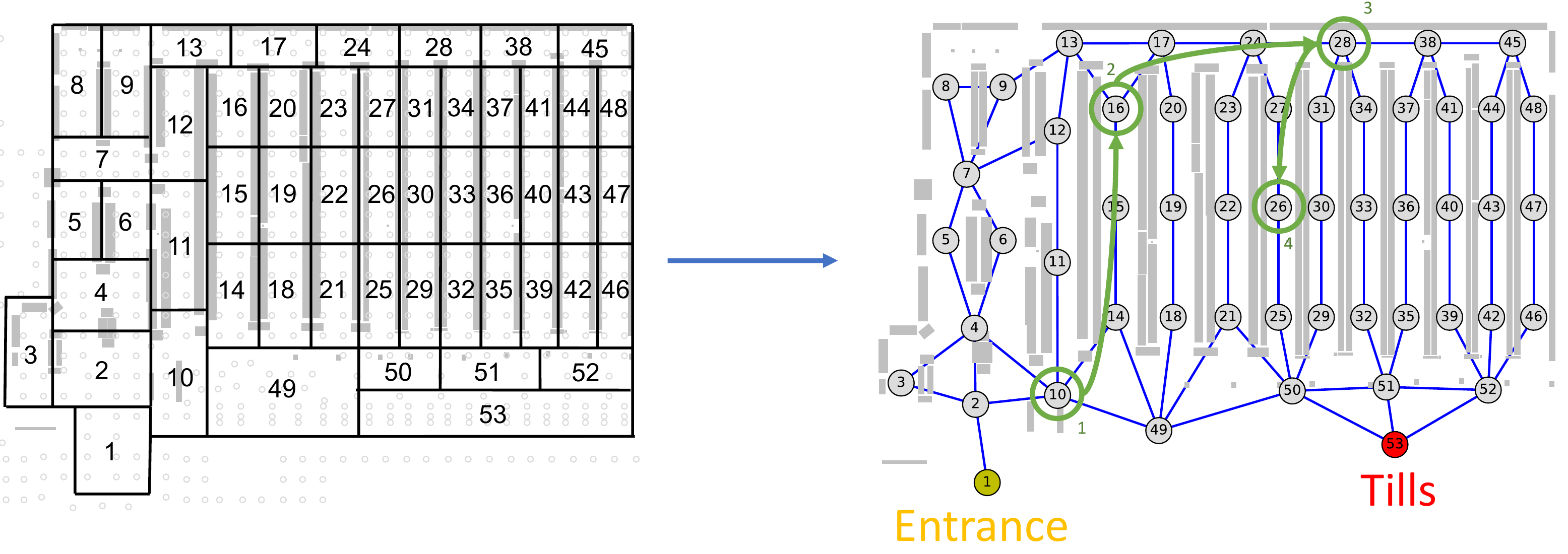}
(b) \includegraphics[width = 1.0\textwidth]{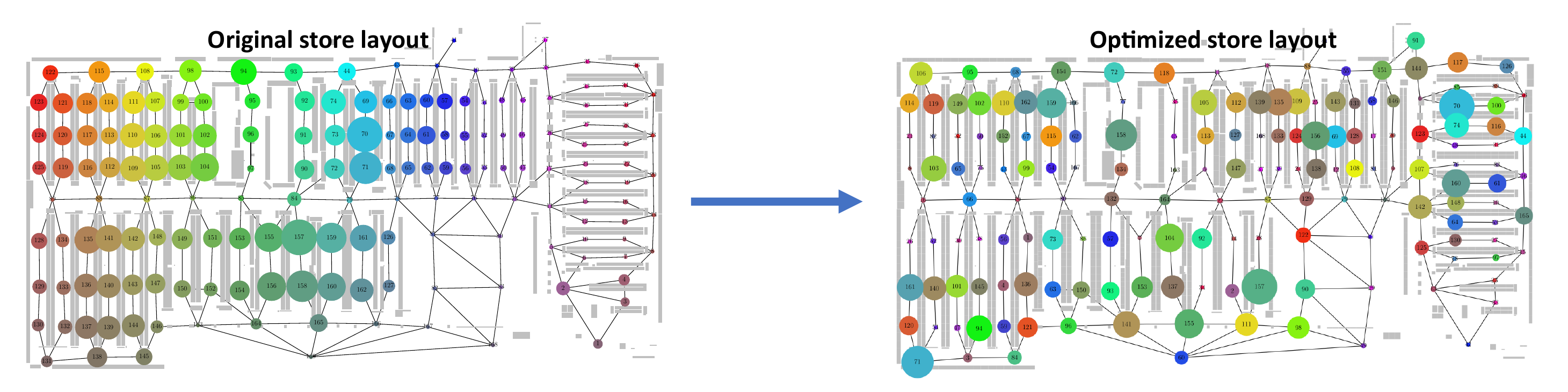}
\caption{(a) A Tesco market (left) is divided into zones, which are the nodes in a network (right). We calculate mobility flow between zones from anonymized shopping-journey data. We show an example customer journey in green.
(b) Original (left) and optimized (right) layouts of a story. In our optimization, we seek to minimize a congestion model. In this optimized store layout, popular zones are on the outer perimeter of the store. Larger nodes have more shopping trips going through them. We use node color to signify positions in the original store layout and to help visualize how they disperse in the optimized layout. [These figures were created by Fabian Ying.]
\label{fabian}
}
\end{figure}

For Fabian's research, we model each store as a
network whose nodes are different regions of that store, with edges present
whenever two regions are next to each other (see Fig.~\ref{fabian}a). There are also specific entrance
and exit points (the tills), and these networks are directed. The spatial
nature of a supermarket network is also an important consideration, as embeddedness in space of a network induces structural features that make them different from other networks \cite{barthelemy2018}. An exciting aspect of Fabian's project is that we are revisiting classical results from stochastic modeling (e.g., queuing theory), dusting them off, and adapting
them for the network age.

Fabian's research involves two interrelated projects. In one project, we seek to minimize congestion in single-source, single-sink queuing networks on which random walkers \cite{masuda2016} (which, for simplicity, we take initially to be homogeneous and unbiased) are traversing. This project, while mathematical in nature, has the potential to give insights on customer congestion in markets. We hope to learn how network structure affects congestion in this model, which network structures minimize congestion (e.g., as measured by mean queue size or the total number of customers who are currently waiting), and how to alter an existing network (e.g., by adding or removing a small number pathways between shopping aisles) to decrease congestion. In the other project, Fabian is using human-mobility models \cite{mobility-review} to analyze the flow of customers between zones in a store. This project involves comparisons of several generative models and a systematic comparison with anonymized data of customer journeys from Tesco stores.

Our approach for the human-mobility project is to study generative models for
customer movements within a store. We are using insights from existing models
for human mobility, which have been a major research topic during the last
decade thanks to the recent availability and abundance of human and animal
mobility data \cite{marta2016,mobility-review}. However, our project entails analysis of rather different temporal and spatial scales
from existing studies. Ultimately, this may necessitate the development of new models, though thus far we have found that existing population-level human mobility models can successfully {\it predict} (not merely fit) about 65--70\% of the mobility flow of customers between zones in a store, which as far as we are aware is the first successful application of these models on such a small spatial scale.
Additionally, through optimization (using simulated annealing), we are able to produce store layouts with less congestion (according to our model), such as by fixing the basic store geometry and swapping zones (see Fig.~\ref{fabian}b). The best layouts bring customers as quickly as possible to a store's exit.

There are numerous exciting directions, which intermingle theoretical and practical foci, to take Fabian's research. For example, it will be useful to study a more realistic routing model between purchases as well as more realistic measures and models for congestion. 
To compare the results of such expanded models with empirical data, we will need to use data sets (e.g., which combine customer-location data and purchase data) that are not currently available to us. It is also important to try to validate our model on a larger number of stores. We also aim to incorporate business constraints into our optimization of store layouts, so that we can provide Tesco with suggestions of viable store layouts.
On the conceptual side, we are fascinated by the idea of augmenting random-walker dynamics through incorporation of shopping lists or zones with different attraction levels, develop new human-mobility models that are tailored for this application's small spatial scale and temporal scales, and so on. 
Potential future avenues include incorporating insights from
recommender systems (e.g., using mobility and behavioral economics) to
influence movement and avoid congestion.


\section{Conclusions} \label{conc}

Network science is playing a large --- and increasing --- role in industrial
problems. Many problems, and associated data, have a natural network
structure; and the study of networks and other discrete structures is
rapidly becoming a core area of applied mathematics alongside traditional
continuum approaches \cite{ejam2016}. The traditional and very successful `physical-applied-mathematics' philosophy
is just as relevant in network modeling as it is in more traditional
applied-mathematics topics, but there are also many fascinating, important,
and often rather difficult challenges: (1) the field of network science is
much less mature than topics such as partial differential equations and
asymptotic analysis, and this necessitates both the development of new methodologies from
industrial (and other application-oriented) problems and the navigation of a situation with less clarity in
how and what level of description to use to attack those problems; (2) because
networks are high-dimensional and the interactions between many entities play
a prominent role in network analysis, modelers should become comfortable not only with traditional mechanistic modeling, a longstanding expertise of applied mathematicians, but also with ideas such as probabilistic modeling, statistics, and
uncertainty quantification; (3) the large scale of networked systems poses
challenges for scientific computation, especially given not only large static
data sets but also real-time computations with data streams; (4) missing and
incomplete data (and data cleaning) provide significant guidance (and
limitations!) that affect not only what calculations are reasonable but also the level of detail that one may wish to use in a network description; and (5) the vast and increasing use of human data poses significant ethical issues (e.g., data privacy) that departs rather markedly from, say, the use of data that arises from fluid mixing in chocolate and other traditional industrial applications
that motivate mathematical studies.

The study of networks is a core part --- and, we would argue, one of the most
important parts --- of the mathematics for the modern economy, as it relates
very strongly both to the types of problems and to the types of data that
arise in it. Network modeling also has deep connections to data analysis,
data science, and Big Data, but it is more than that: network science
incorporates modeling tenets from both physical applied mathematics and
statistics, and it seeks to marry them together. Mathematics departments need
to develop strength in network modeling, and those efforts need to include
the study of problems with close ties to problems that arise in the
industrial, commercial, and governmental sectors. This will help solve
important societal problems and simultaneously lead to the develop of new
mathematical and computational techniques and insights.


\section*{Acknowledgements}

We thank John Ockendon for the invitation to write this article and to give an associated talk at the Royal Society workshop on `Mathematics for the Modern Economy' \cite{roysocmeeting}.

We thank Robert Armstrong, Mariano Beguerisse D\'iaz, Jeremy Bradley, Peter Grindrod, Valid Krebs, Ursula Martin, Roxana Pamfil, Rosie Prior, and Fabian Ying for helpful comments. We also thank Roxana Pamfil for creating the original versions of Figs.~\ref{bipartite} and \ref{complicated} and helping us modify them, Fabian Ying for creating Fig.~\ref{fabian}, and Andrew Stuart for an inspiring slide and discussion that we adapted for one of our paragraphs. We thank several anonymous referees for helpful comments.

Our doctoral students Fabian Ying and Roxana Pamfil were funded by the EPSRC Centre For Doctoral Training in Industrially Focused Mathematical Modelling (EP/L015803/1) in collaboration with Tesco and dunnhumby, respectively.  Our doctoral students Daniel Fenn and Marya Bazzi were funded by the EPSRC through CASE studentship awards.







\end{document}